\documentclass[11pt,a4paper]{article}

\usepackage[margin=1in]{geometry}
\usepackage[T1]{fontenc}
\usepackage[utf8]{inputenc}
\usepackage{lmodern}
\usepackage{microtype}
\usepackage{parskip}
\usepackage{placeins}
\usepackage{amsmath,amssymb,amsfonts,bm}
\usepackage{physics}

\usepackage{siunitx}
\usepackage{graphicx}
\usepackage{subcaption}
\usepackage{booktabs}
\usepackage{multirow}
\usepackage{array}

\usepackage{authblk}

\usepackage[numbers,sort&compress]{natbib}
\usepackage{hyperref}
\usepackage[nameinlink,capitalise,noabbrev]{cleveref}
\hypersetup{
  colorlinks=true,
  linkcolor=blue,
  citecolor=blue,
  urlcolor=blue,
  pdftitle={Continuous-Wave SOFISM with SPAD Array Detection},
  pdfauthor={LM Beck, T Dadosh, A Chen, I Goliand, D Oron},
}

\newcommand{\sofiism}{SOFISM}

\newcommand{\figref}[1]{\cref{#1}}

\renewenvironment{abstract}
 {\small
  \begin{center}
  \bfseries Abstract\vspace{-.5em}\vspace{0pt}
  \end{center}
  \list{}{
    \setlength{\leftmargin}{1.5cm}%
    \setlength{\rightmargin}{\leftmargin}%
  }%
  \item\relax}
 {\endlist}

\title{Continuous-Wave SOFISM with SPAD Array Detection}

\author[1,4]{Lior M. Beck}
\author[2]{Tali Dadosh}
\author[1]{Anabel Chen}
\author[3]{Inna Goliand}
\author[1,5]{Dan Oron}

\affil[1]{Department of Molecular Chemistry and Materials Science, Weizmann Institute of Science, Rehovot 7610001, Israel}
\affil[2]{Department of Chemical Research Support, Weizmann Institute of Science, Rehovot 7610001, Israel}
\affil[3]{Department of Life Sciences Core Facilities, Weizmann Institute of Science, Rehovot 7610001, Israel}
\affil[4]{\texttt{lior.beck@weizmann.ac.il}}
\affil[5]{\texttt{dan.oron@weizmann.ac.il}}

\date{}

\begin{document}
\maketitle

\begin{abstract}
We present a continuous-wave (CW) implementation of super-resolution optical fluctuation imaging with image scanning microscopy (SOFISM) using a pixelated single-photon avalanche diode (SPAD) detector camera. In a scanning-based geometry, SOFISM requires fluorescence fluctuations on timescales compatible with the pixel dwell time while maintaining sufficient signal under confocal excitation. Photoswitchable fluorophores in an aqueous switching buffer containing the glucose oxidase-catalase oxygen-scavenging system (GLOXY) and mercaptoethylamine (MEA) were therefore used to promote blinking and reduce photobleaching. Correlation analysis of sparse-emitter measurements showed that Alexa Fluor 647 (AF647) exhibits microsecond-scale fluctuation dynamics suitable for CW SOFISM and guided the choice of imaging conditions for scanned samples. Frame rate and delay accumulation were optimized from detector-pair subset variability using a mean-to-STD criterion. The optimal binning time was \SI{2.5}{\micro\second}, underscoring the advantage of SPAD-based detection for resolving the relevant fluctuation dynamics. The method was demonstrated on HeLa cell microtubules labeled with AF647, yielding improved spatial resolution in the constructed SOFISM images compared with the corresponding ISM images. Additional super-resolved images from samples labeled with CF568 and Alexa Fluor 555 (AF555) showed the approach is readily generalizable to other fluorophores. These results establish CW SOFISM as a practical fluctuation-based super-resolution method for confocal microscopy.
\end{abstract}

\section{Introduction}
\label{sec:intro}


The resolution of any optical imaging apparatus is limited by the Abbe diffraction limit, corresponding to the wavelength divided by twice the numerical aperture~\cite{abbe1873beitrage}. This limit relies on specific physical assumptions. By breaking these assumptions, such as the wave nature of light, the linearity of sample response, and sample stationarity, it is possible to further improve spatial resolution.

Early super-resolution strategies achieved this through single-molecule localization or targeted depletion. These strategies are seen in methods like STORM, PALM, and STED~\cite{rustSubdiffractionlimitImagingStochastic2006, betzigImagingIntracellularFluorescent2006b, hellBreakingDiffractionResolution1994a}. Stochastic Optical Reconstruction Microscopy (STORM), in particular, relies on the ability to switch fluorescent emitters between a bright "on" state and a dark "off" state. This requires both special fluorophores and a special environment, such as a STORM buffer containing reducing agents and oxygen scavengers~\cite{dempseyEvaluationFluorophoresOptimal2011a}. These buffers drive compatible fluorophores into long-lived dark states while simultaneously protecting the sample from oxidative damage. This protection significantly reduces photobleaching and improves the total photon budget~\cite{aitkenOxygenScavengingSystem2008}, allowing for the precise localization of individual emitters over time. Such buffer strategies are also broadly relevant in other super-resolution imaging techniques, where balancing photostability with controlled emitter dynamics is essential for robust image reconstruction.

While these localization-based techniques provide exceptional precision, they often require specific chemical environments to facilitate emitter switching. Other super-resolution modalities have been developed that circumvent the diffraction limit by manipulating illumination patterns or detection geometry rather than single-molecule switching. Structured Illumination Microscopy (SIM), for instance, extends the measurable spatial bandwidth by using patterned widefield illumination to sample otherwise inaccessible spatial frequencies~\cite{heintzmannLaterallyModulatedExcitation1999a}. Related approaches such as Random Illumination Microscopy (RIM) use unknown or random illumination patterns to achieve similar bandwidth extension~\cite{mangeatSuperresolvedLivecellImaging2021}. Its counterpart, Image Scanning Microscopy (ISM), samples the spatial frequencies through confocal scanning and constructs the super-resolved image by pixel reassignment (PR) of the collected detector images~\cite{sheppardSuperresolutionConfocalImaging1988, mullerImageScanningMicroscopy2010}.

Further improvements can be made by exploiting the statistical or quantum properties of the emitted light. Super-Resolution Optical Fluctuation Imaging (SOFI) utilizes fluorescence fluctuations similar to those used in STORM, but it processes them through higher-order cumulants rather than individual localization~\cite{dertingerFastBackgroundfree3D2009}. Because these emitters blink independently, cross-correlations between different emitters tend to vanish, while higher-order cumulants preserve information specific to individual emitters. For the independent blinking emitters, the $n$th-order cumulant contains an effective PSF proportional to the $n$th power of the optical PSF, narrowing the response by approximately $\sqrt{n}$ for a Gaussian PSF. A related modality, Quantum Image Scanning Microscopy (Q-ISM), relies on the quantum nature of light rather than stochastic blinking. This method utilizes photon antibunching, where the "missing" simultaneous photon pair events provide a contrast mechanism that yields, similarly to SOFI, a narrower PSF~\cite{schwartzSuperresolutionMicroscopyQuantum2013}.

It is also possible to combine these modalities to leverage their respective strengths. Examples include SOFI-SIM~\cite{desclouxExperimentalCombinationSuperResolution2021b}, which integrates fluctuation analysis with structured widefield patterns to further enhance resolution and optical sectioning, and Q-ISM, combining quantum fluctuations with an ISM setup~\cite{tenneSuperresolutionEnhancementQuantum2019}.

Another modality, and the focus of this paper, is the combination of SOFI and ISM, termed SOFISM~\cite{srodaSOFISMSuperresolutionOptical2020b}. This technique uses confocal scanning while fluorescence fluctuations are recorded at each scan position with a pixelated detector. SOFISM images are formed from correlation channels between detector pairs. Each detector pair can be viewed as a virtual detector located between the two physical detector elements. PR then shifts each correlation channel according to this detector-pair geometry, and the reassigned channels are summed to form the final image. The reassignment and summation step improves spatial resolution while averaging over multiple detector-pair realizations, which helps stabilize the fluctuation contrast.

Modern Single-Photon Avalanche Diode (SPAD) array technology has been a primary enabler for this modality. These cameras offer essentially zero readout noise, sub-nanosecond timing accuracy, and saturation fluences of tens of MHz. This performance allows for microsecond-scale binning, which provides a sufficient number of frames within each pixel dwell time to perform SOFISM processing effectively.

However, the fluctuation signal that drives SOFI contrast is inherently noisy. In a scanning ISM configuration, this SNR limitation is made more severe by the limited photon budget and short dwell time per pixel. To address this, several solutions have been proposed. These include sparse-constraint reconstructions~\cite{rossmanRapidQuantumImage2019}, and an image fusion method
driven by self-supervised deep learning ~\cite{beckImprovingCorrelationBased2024}. 
Photobleaching presents an additional hurdle, as it further reduces an already sparse signal and can introduce additional noise into the cumulant estimation.

Despite these challenges, the motivation for refining SOFISM is clear. ISM is already available in commercial systems such as the Zeiss Airyscan, while SPAD-array-based ISM detection is implemented in platforms such as Nikon's NSPARC~\cite{huffAiryscanDetectorZEISS2015,delattreIgnitingNewConfocal2023}. These developments suggest that methods compatible with SPAD-based ISM architectures may become increasingly relevant as commercial implementations continue to evolve. It is important to note that these commercial units typically operate using continuous-wave (CW) illumination. This stands in contrast to previous implementations of SOFISM, which utilized pulsed-laser configurations to specifically capture triplet-state dynamics~\cite{srodaSOFISMSuperresolutionOptical2020,krupinski-ptaszekSuperresolutionMicroscopyBased2025}. Therefore, a modality that increases resolution under standard CW conditions, without requiring hardware changes to these widely deployed setups, would be highly practical for the broader research community.

In this paper, we present a practical implementation of SOFISM optimized for CW illumination using a custom galvo-scanning system and a SPAD detector array. We first validate the correlation signal using sparse emitters to establish a baseline for imaging feasibility and signal-to-noise ratio (SNR) performance. We then apply this framework to HeLa cell samples to examine the interplay between scanning parameters and fluorescence kinetics, specifically characterizing how different buffer environments influence the resulting super-resolved images. Finally, we demonstrate the versatility of this approach by imaging multiple common fluorophores, including Alexa Fluor 647 (AF647), Alexa Fluor 555 (AF555) and CF568.

\section{Methods}
\label{sec:methods}
\subsection*{Optical setup and data acquisition}
\begin{figure}[!b]
    \centering
    \includegraphics[width=0.9\linewidth]{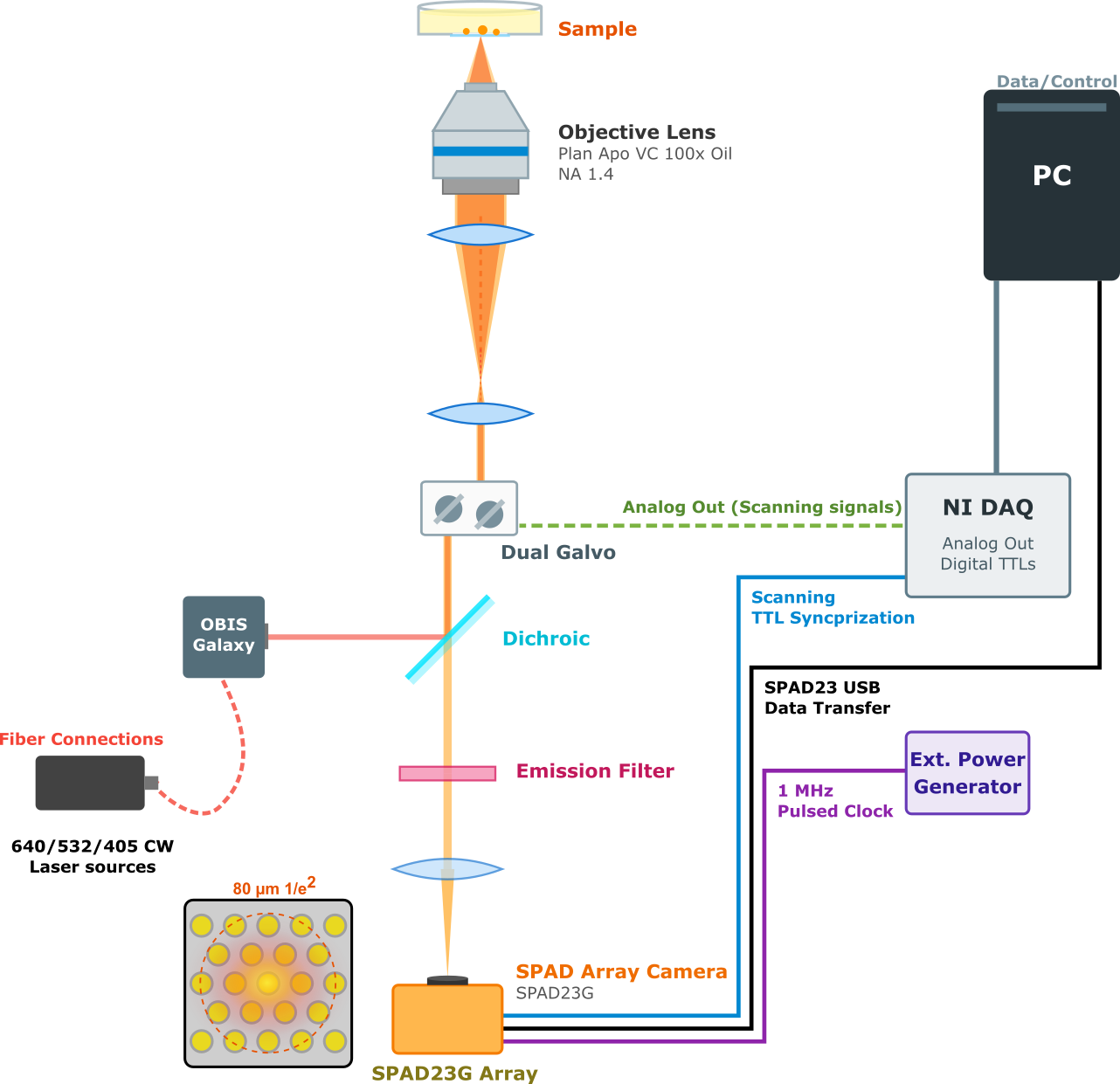}
    \caption{Experimental setup schematic for CW-\sofiism{} imaging. A 640/532~nm CW excitation beam is scanned by galvo mirrors across the sample in a confocal geometry. The emitted fluorescence is descanned and projected onto a 23-pixel SPAD array, with magnification adjusted to match the Gaussian to the detector size.}
    \label{fig:setup}
\end{figure}

The experimental setup is based on a custom confocal microscope built around the body of an altered Axiovert 135 (Zeiss)(\figref{fig:setup}).

The sample was excited primarily with a 640~nm continuous-wave (CW) laser (OBIS 640~nm LX 75~mW, Coherent). For a separate set of samples, excitation was instead provided by a 532~nm CW laser (OBIS 532~nm LS 80~mW, Coherent). An additional 405~nm laser source (OBIS 405~nm LX 50mW, fiber pigtail, UFC, Galaxy) was included in the system and used mainly for alignment. The excitation beams were fiber-coupled into a combiner unit (OBIS Galaxy, Coherent). Beam scanning was performed with a dual-axis galvo scanner (GVS012, Thorlabs), while excitation and detection were separated by a dichroic mirror (di03-r405/488/532/635, Semrock). The beam then entered the scanning arm of the microscope, where the dual-axis galvo was followed by a 2$f$ relay composed of a scan lens and a tube lens, which filled the back aperture of the objective and produced a diffraction-limited excitation spot at the sample plane.

Fluorescence was collected in epi-detection configuration by an oil-immersion objective of numerical aperture (NA) 1.4 (Plan Apo VC 100$\times$ Oil $\infty$/0.17 DIC N2, Nikon). The emitted signal was descanned through the galvo mirrors and directed to the detection path, where residual excitation light was rejected with an emission filter (ZET532/640m, Chroma). The fluorescence signal was then focused onto the SPAD array camera (SPAD23G, Pi Imaging). The detector pixels are hexagonally packed, with each SPAD detector having a diameter of \SI{20}{\micro\meter}. Using a fluorescent bead, a variable telescope was adjusted such that the diffraction-limited spot spanned the array. The resulting illumination spot on the SPAD array was fitted with a two-dimensional Gaussian, yielding a $1/e^2$ diameter of approximately \SI{80}{\micro\meter}, thus filling most of the detector (leaving out the corner detectors).

The motion of the galvo mirrors is controlled by the analog output of a data-acquisition (DAQ) unit (USB-6341, NI) connected to a PC.
Two digital channels in the DAQ generate transistor-transistor logic (TTL) pulses for the SPAD camera to synchronize the lateral steps during the scan.
The SPAD camera was operated in timestamp mode using an external
1 MHz reference clock.

\subsection*{Temporal binning, filtering, and correlation analysis}
The raw time-stamped photon detections from the SPAD array were binned into a sparse intensity matrix with a temporal resolution of \(\Delta t = \SI{2.5}{\micro\second}\) per scan position. The edges of each pixel dwell time were trimmed to remove galvanometer mirror settling times (a few hundred \si{\micro\second}).

Periodic intensity modulations at harmonics of 100\,Hz, attributed to power-line interference coupling into the galvanometer drive, were suppressed by applying a sigmoid high-pass filter at $f_c = 250$\,Hz in the frequency domain. After filtering, the fluctuation temporal cross-correlation between all detector pairs $(i,j)$ was computed for each scan position $\mathbf{r}$ and delay $\tau$:
\begin{equation}
    C_{ij}(\mathbf{r}, \tau) = \frac{1}{N_t - \tau} \sum_{t=1}^{N_t - \tau} \delta I_i(\mathbf{r}, t)\,\delta I_j(\mathbf{r}, t+\tau)
    \label{eq:cross_corr}
\end{equation}
where \(\delta I_i(\mathbf{r},t)\) is the high-pass-filtered fluctuation signal of detector \(i\) at scan position \(\mathbf{r}\), and \(N_t\) is the number of temporal bins after trimming. For the static sparse-emitter measurements, the continuous timestamp trace was additionally segmented into shorter blocks, which were treated as independent realizations before averaging the correlations.

\subsection*{Pixel reassignment and SOFISM image formation}
To perform pixel reassignment and produce an ISM image, a detector-resolved Confocal Laser Scanning Microscopy (CLSM) image was generated for each SPAD element. Each detector image was compared with the central detector image using sub-pixel cross-correlation of the image gradients~\cite{guizar-sicairosEfficientSubpixelImage2008}. The displacement that minimized the gradient-image mismatch was used as the pixel-reassignment shift $\vec{\nu}_i$. This data-driven procedure avoids the need for a separate calibration measurement and compensates for small alignment changes between experiments.

The same displacement vectors were used to correct the detector-pair parallax in SOFISM. For detector pair $(i,j)$, the effective shift was taken as the average of the two individual detector shifts, since the signal originates physically from the midpoint between the detectors:
\begin{equation}
    \vec{\nu}_{ij} = \frac{\vec{\nu}_i + \vec{\nu}_j}{2}
    \label{eq:sofism_shift}
\end{equation}
Each slice of the correlation stack was translated by $\vec{\nu}_{ij}$, and the shifted images were summed over all cross-correlation pairs ($i \neq j$). Autocorrelation terms, where $i=j$, were ignored, as explained further in the Results section.

\subsection*{Fourier reweighting}
To obtain the final super-resolved image, the summed SOFISM image was Fourier-reweighted (FR) using a Wiener-type deconvolution filter constructed from the squared ISM optical transfer function (OTF) measured on a calibration bead:
\begin{equation}
    W(\mathbf{k}) = \frac{1}{\mathrm{OTF}_{\mathrm{ISM}}^2(\mathbf{k}) + \varepsilon}
    \label{eq:fourier_reweight}
\end{equation}
where $\mathrm{OTF}_{\mathrm{ISM}}(\mathbf{k})$ is the measured ISM optical transfer function and $\varepsilon$ is a regularization parameter.
Further method description can be found in~\cref{sec:si_sofism_pipeline}.

\subsection*{Samples and imaging buffers}

Sparse AF647 samples were prepared by nonspecific adsorption of AF647-labeled
secondary antibodies onto poly-L-lysine-coated glass. These samples were used
for static fluctuation measurements under CW excitation. For cellular imaging, HeLa cells were fixed, permeabilized, and immunostained against $\alpha$-tubulin. Measurements were performed using AF647, CF568 or
AF555 labeled secondary antibodies, as indicated in the relevant experiments. Imaging was performed in glucose-containing Tris buffer supplemented with
glucose oxidase/catalase and mercaptoethylamine. These
buffers were used to promote fluorescence fluctuations while reducing
photobleaching during repeated scanning. Additional details regarding sample preparation and buffer protocols are provided in~\cref{sec:si_sample_prep_buffers}.

\section{Results}
\label{sec:results}
\subsection{Fluorophore blinking requirements for scanning SOFISM}

In SOFISM, the confocal scan poses a limit both on the amount of signal that can be acquired within a single dwell time and the amount of fluctuation each emitter possesses in that time to give enough contrast to construct a super-resolved image. This is in contrast to camera-based SOFI realizations, which typically require very long (> 1~ms) on/off times, limited by the camera frame rate.

To recover the necessary blinking kinetics and mitigate photobleaching, we used an aqueous imaging buffer with AF647 as the primary fluorophore, since it is well established for SOFI, and STORM-based imaging~\cite{dertingerSuperresolutionOpticalFluctuation2010,karlssonPhotoisomerizationCyanineDye2019}. Additional measurements with AF555 and CF568 were used to test whether the approach could be extended to other labels. In all cases, application to a scanning-based SOFISM architecture requires careful examination of the characteristic fluctuation time. The stochastic blinking must be sufficiently fast to be captured within the limited pixel dwell time inherent to a scanning configuration, which utilizes a completely different timescale dynamics than STORM or widefield SOFI implementations~\cite{krupinski-ptaszekSuperresolutionMicroscopyBased2025}.

\subsection{AF647 sparse-emitter dynamics}

\begin{figure}[!htbp]
    \centering
    \includegraphics[width=\linewidth]{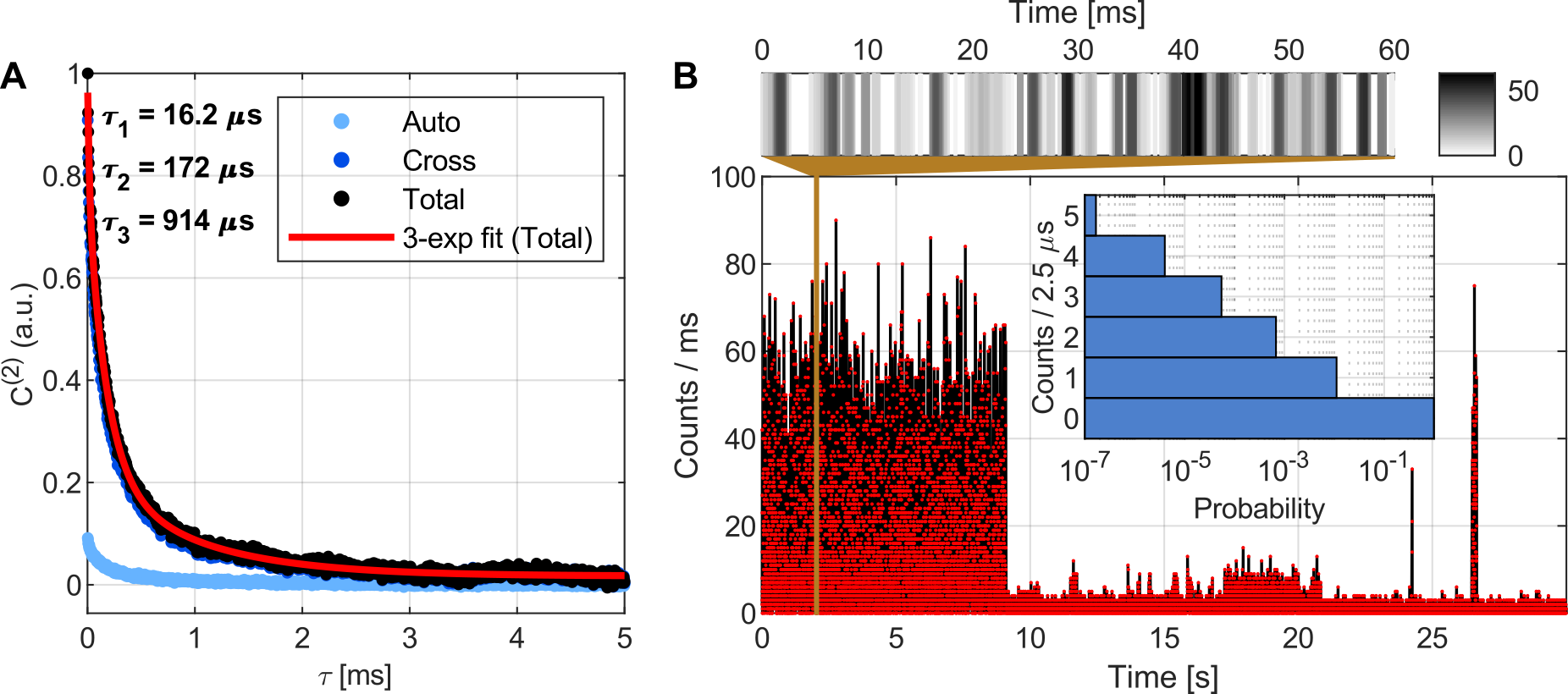}
    \caption{Static sparse-emitter AF647 measurement, acquired with a native bin time of \SI{2.5}{\micro\second}. \textbf{(a)} Sum of auto- and cross-correlation (and each one separately) fluctuation signals over a 5~ms delay range, fit with a tri-exponential decay (red). The three extracted time constants ($\tau_1$, $\tau_2$, $\tau_3$). \textbf{(b)} \textit{Bottom:} full 30~s photon-count trace, rebinned to 1~ms for visualization. \textit{Inset:} counts-per-bin probability histogram at the native \SI{2.5}{\micro\second} binning (log-scale probability). \textit{Top:} density-coded stripe showing each photon-containing \SI{2.5}{\micro\second} bin within the highlighted 60~ms window (orange) of the bottom trace; line color encodes the local count rate measured within a 1~ms sliding window centered on each bin.}
    \label{fig:fig2_trace_corr}
\end{figure}

Prolonged exposure to high continuous-wave (CW) intensities heavily populates the triplet state, trapping fluorophores in long-lived dark states that suppress fast blinking kinetics and severely exacerbate photobleaching~\cite{byrdinImpactTripletState2025}. Therefore, these measurements require a medium that limits photobleaching while preserving the fluorescence fluctuations. 
The buffer system consisted of tris buffer at pH 8.5 with an the oxygen-scavenging system glucose oxidase-catalase (GLOXY), and the reducing agent mercaptoethylamine (MEA).
The measurement consists of 30 seconds of timestamped data. The data was binned to  \SI{2.5}{\micro\second} and segmented into 500 short traces of 60~ms duration to approximate the limited acquisition window of the confocal implementation. Reference samples in phosphate-buffered saline (PBS) photobleached rapidly and yielded little to no usable fluctuation-correlation signal.
Measurements were acquired at excitation powers of \SI{100}{\nano\watt}, \SI{1}{\micro\watt}, and \SI{2}{\micro\watt}. For \SI{640}{\nano\meter} illumination, this corresponds to approximately
\SI{0.08}{\kilo\watt\per\square\centi\meter},
\SI{0.8}{\kilo\watt\per\square\centi\meter}, and
\SI{1.6}{\kilo\watt\per\square\centi\meter}.
AF647 produced the strongest correlation signal at \SI{2}{\micro\watt} excitation in an imaging buffer containing GLOXY and MEA. An example of such a static measurement is shown in Fig.~\ref{fig:fig2_trace_corr}. A tri-exponential fit revealed multiple blinking timescales, with characteristic components of approximately \SI{16}{\micro\second}, \SI{170}{\micro\second}, and \SI{915}{\micro\second}.

\FloatBarrier
\subsection{AF647 correlations under confocal scanning}

Microtubules in HeLa cells immunolabeled with AF647 were imaged to evaluate CW SOFISM performance under different buffer and laser-power conditions. These measurements test whether the fluctuation behavior identified in sparse-emitter experiments could be extended to biological samples under confocal imaging conditions.

Because the confocal measurements contained a pronounced low-frequency component at 100~Hz from the galvo mirrors, the intensity traces were temporally high-pass filtered before correlation analysis. This filtering step was necessary to isolate the fluctuation components that could be used reliably for SOFISM image construction.

The fitted decay times varied only modestly between buffer compositions, as summarized in Table~\ref{tab:buffer_correlation}. After filtering, the apparent time constants were slightly shorter than those extracted from the unfiltered sparse-emitter data, consistent with the preferential suppression of slow fluctuations by the high-pass filter. This comparison, shown in Fig.~\ref{fig:fig2_trace_corr}(a) and Fig.~\ref{fig:fig3_staticVsScan0}(a), provides an estimate of the fluctuation timescales that remain accessible for image formation under the filtered scanned-imaging conditions.

The autocorrelation terms became negative at a nonzero delay after filtering, as shown in Fig.~\ref{fig:fig3_staticVsScan0}(b). This behavior is expected after high-pass filtering. The filter constrains the total area under both autocorrelation and cross-correlation curves to be close to zero. For autocorrelation the zero-delay contains a large shot-noise contribution. This contribution must be compensated by negative values at nonzero delays. The resulting nonzero-delay autocorrelation is therefore strongly shaped by the filtering constraint and shot noise, rather than by fluorophore fluctuation dynamics. For this reason, the final contrast for SOFISM images was generated using only detector cross-correlations.

Taken together, these measurements showed that the samples were robust over the tested buffer compositions when both MEA and GLOXY were present. In contrast, when MEA was omitted, the fluctuation signal depended approximately linearly on excitation power, suggesting that the reducing buffer component plays an important role in producing the nonlinear blinking response required for SOFISM. This power dependence is examined further in~\cref{sec:si_buffer_behavior} and summarized in~\cref{fig:si_tau_boxplot}.

\begin{table}[htbp]
\centering
\caption{Bi-exponential correlation fit parameters for scanned samples under different imaging-buffer compositions. MEA and GLOXY are reported as stock-solution fractions in the final 1~ml imaging buffer.}
\label{tab:buffer_correlation}
\begin{tabular}{cccc}
\hline
MEA stock (\% v/v) & GLOXY stock (\% v/v) & $\tau_\mathrm{fast}$ (\si{\micro\second}) & $\tau_\mathrm{slow}$ (\si{\micro\second}) \\
\hline
2 & 5 & $8.4 \pm 1.0$ & $92.4 \pm 10.4$ \\
5 & 2 & $7.4 \pm 1.4$ & $88.5 \pm 22.9$ \\
0 & 5 & $8.8 \pm 2.0$ & $99.0 \pm 13.4$ \\
\hline
\end{tabular}
\end{table}

The confocal dataset shown in Fig.~\ref{fig:fig3_staticVsScan0} was acquired with a dwell time of 10~ms, a step size of 50~nm, five repeats, and an excitation power of \SI{3}{\micro\watt}. Comparison of the photon counts at a bright pixel across repeats indicates that photobleaching is still present. Accordingly, the correlation amplitude gradually decreases from repeat to repeat, reducing the contribution of later repeats to the final SOFISM image contrast. 

\subsection{Temporal binning and delay selection}

To select the temporal binning and correlation-delay range used for SOFISM image formation, an internal detector-pair stability metric was computed directly from the measured cross-correlation data. This optimization was required because the useful SOFISM contrast depends on the temporal sampling of the fluorescence fluctuations, whereas summing correlation delays beyond the decay of the fluctuation signal can add noise without improving image contrast. Since SOFI signal is highly non-uniform across the field of view and no ground-truth reference image is available, the metric was based on the consistency of the SOFISM contrast across detector-pair correlations.

For each tested temporal binning and accumulated delay range, the detector cross-correlation terms were divided into signal-balanced groups, with autocorrelation terms excluded. A partial SOFISM contribution image was formed from each group, and a local stability metric was defined as the ratio between the mean group-wise contribution and the standard deviation across groups. This quantity was used as an empirical measure of how consistently the SOFISM contrast was supported by different subsets of detector-pair correlations. The grouping procedure, variance calculation, and repeat-based control analysis are described in~\cref{sec:si_selection_framerate_delay}.

\begin{figure}[!htbp]
    \centering
    \includegraphics[width=0.9\linewidth]{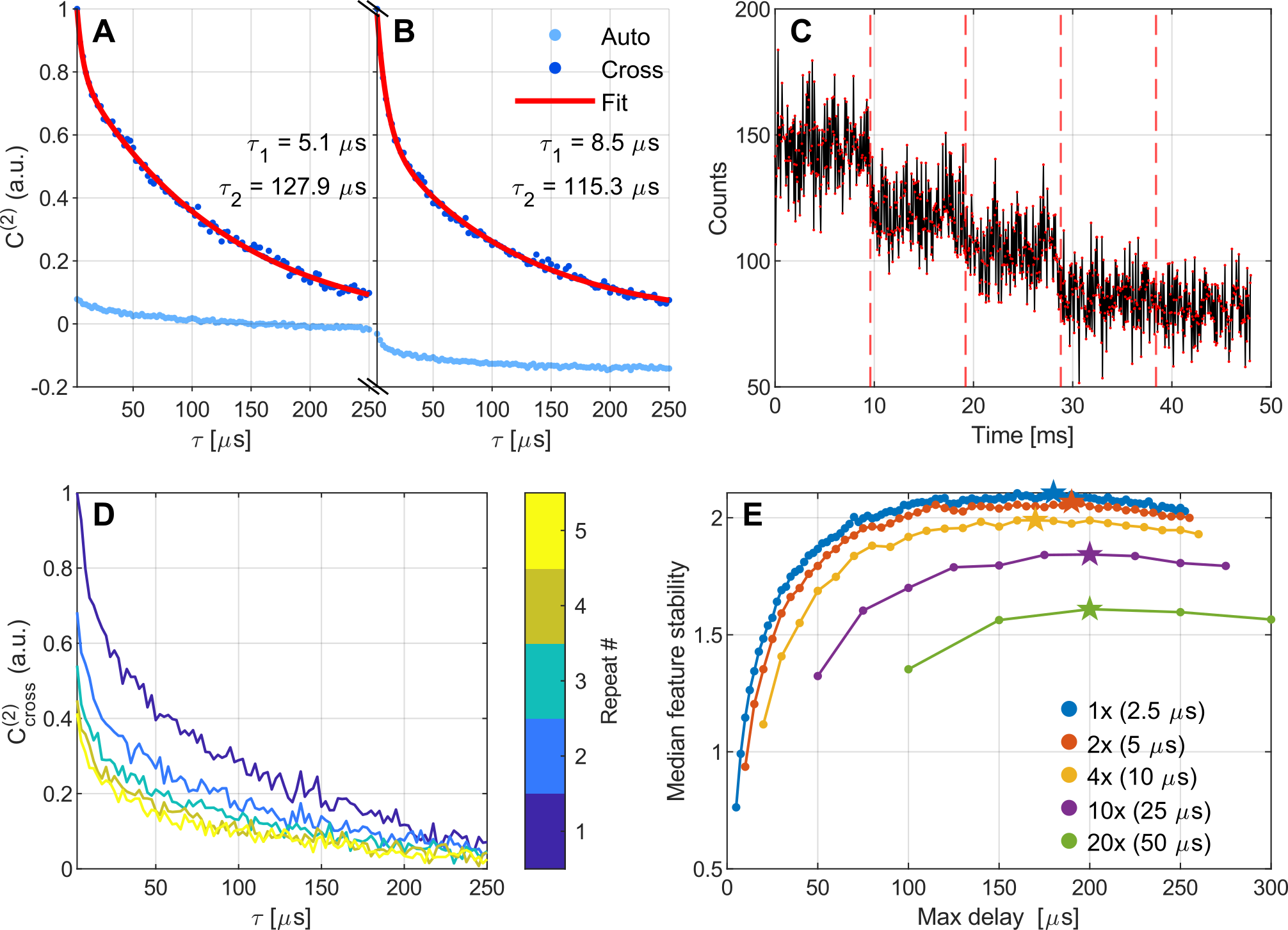}
    \caption{Connecting static emitter dynamics to scanned image correlations.
        (a)~Auto and cross-correlation of the high-pass-filtered static AF647 measurement with a bi-exponential fit to the cross-correlation \(\left(\tau_1 \approx \SI{5}{\micro\second}, \tau_2 \approx \SI{128}{\micro\second}\right)\).
        (b)~Auto and cross-correlation summed over all image pixels and detector pairs, with a bi-exponential fit to the cross-correlation
\(\left(\tau_1 \approx \SI{8.5}{\micro\second}, \tau_2 \approx \SI{115}{\micro\second}\right)\).
        (c)~High-pass-filtered intensity trace of a selected bright pixel, separated by scan repetitions and binned to \SI{50}{\micro\second} resolution ($20\times$ coarse binning). Red dashed lines mark repetition boundaries.
        (d)~Per-repetition cross-correlation curves summed over all image pixels, colored by repetition number.
        (e)~Detector-pair stability metric versus accumulated
        correlation delay for several temporal binnings. Stars mark the
        per-binning optimum.}
    \label{fig:fig3_staticVsScan0}
\end{figure}

The metric was evaluated as a function of temporal binning and maximum accumulated correlation delay. This allowed the frame time to be chosen such that fast fluctuation components were retained, while the accumulated delay range was restricted to delays that contributed useful correlation contrast. The analysis in Fig.~\ref{fig:fig3_staticVsScan0}(e) showed that shorter temporal binning produced stronger usable SOFISM contrast, provided that a sufficient number of correlation delays were accumulated, consistent with the bi-exponential fluctuation dynamics extracted above.

\FloatBarrier
\subsection{Imaging of AF647-Labeled Microtubules in Cells}

\begin{figure}[!htbp]
    \centering
    \includegraphics[width=0.9\linewidth]{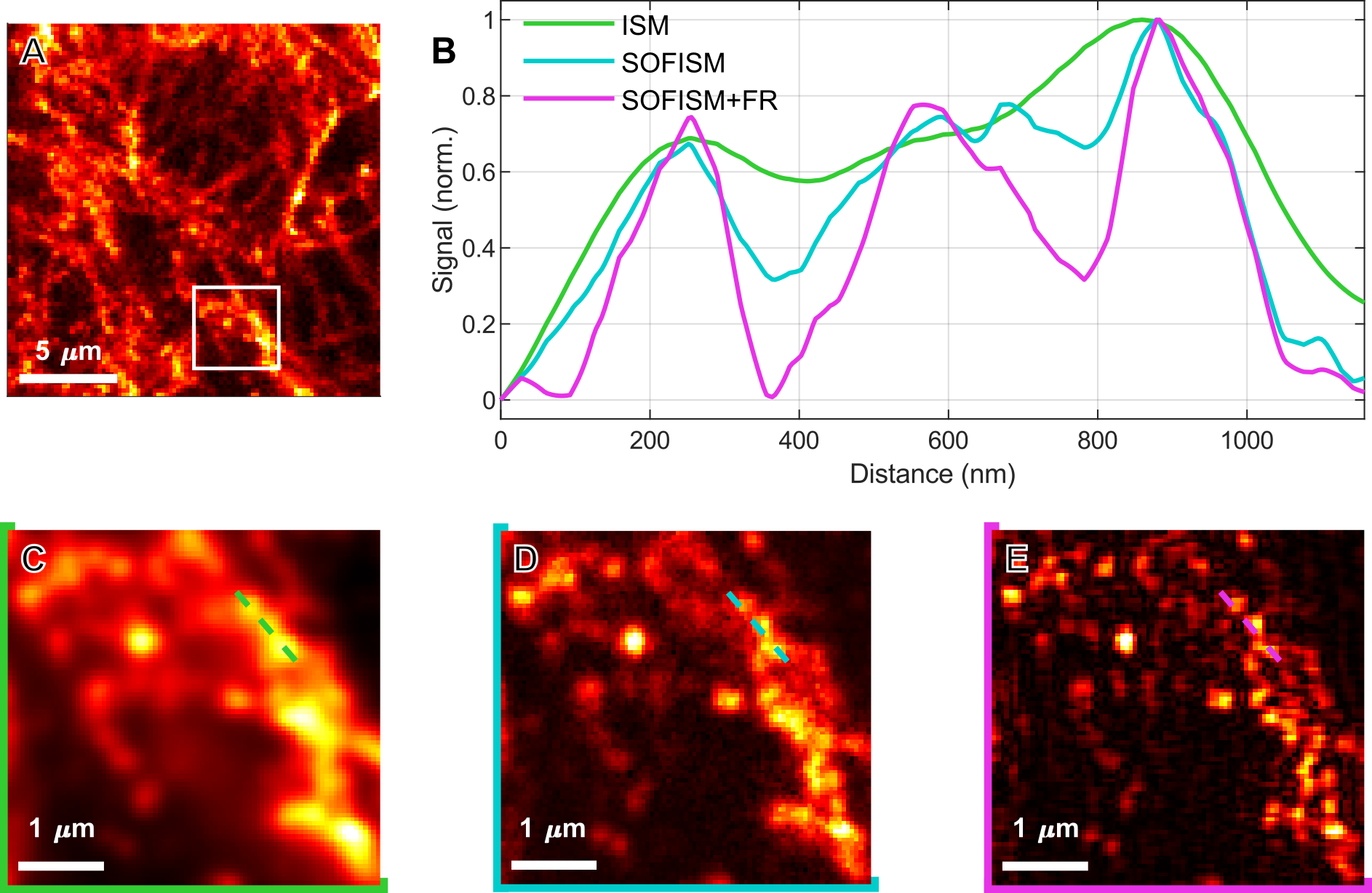}
    \caption{Representative HeLa cell with microtubules labeled with AF647. Image comparison (a)  Large FOV of a fast coarse scan with 1~ms dwell time and 200~nm step size (b) Intensity cross-section comparing ISM, SOFISM, and FR SOFISM. (c,d,e) Different images of the same measurement - 10~ms dwell time, 50~nm step size and 5 repetitions. (c) ISM image. (d) SOFISM image. (e) Fourier-reweighted SOFISM image. }
    \label{fig:fig4_imageComparison0}
\end{figure}

To evaluate SOFISM performance in a biologically relevant sample, we imaged cellular microtubules labeled with AF647. In practice, SOFISM imaging was performed in two stages. First, a large field of view was acquired to identify a suitable region for measurement. As shown in Fig.~\ref{fig:fig4_imageComparison0}, a coarse scan covering a \SI{20}{\micro\meter} \(\times\) \SI{20}{\micro\meter} field of view was recorded using a 200~nm step size, 1~ms dwell time, and a lower average excitation power of \SI{0.1}{\micro\watt} in order to limit photobleaching, this initial scan required only 10~s. After selecting a region of interest, a higher-resolution confocal scan was acquired for SOFISM image. The images shown in Fig.~\ref{fig:fig4_imageComparison0} were recorded with a dwell time of 10~ms, a step size of 50~nm, an excitation power of \SI{3}{\micro\watt}, and with five repeats (same data shown in Fig ~\ref{fig:fig3_staticVsScan0}). To form the SOFISM image, the accumulated delay range was selected from the mean-to-STD analysis in Fig. \ref{fig:fig3_staticVsScan0} (e). This analysis identified \SI{180}{\micro\second} as the delay cutoff that maximized the reproducibility of the SOFISM contrast across the detector-pair subsets.
A normalized cross section taken along the dashed line demonstrates the expected resolution improvement: the SOFISM image resolves features that are not clearly separated in the corresponding ISM image, while the Fourier-reweighted SOFISM image provides a further improvement in feature separation.
The stronger background observed in the ISM image compared with the corresponding SOFISM image is consistent with SOFISM's improved optical sectioning. This improvement originates from the sharper effective axial PSF of the correlation image, which suppresses out-of-focus fluorescence more strongly than the intensity-based ISM image~\cite{srodaSOFISMSuperresolutionOptical2020}.

\FloatBarrier
\subsection{Extension to CF568 and AF555}

\begin{figure}[!htbp]
    \centering
    \includegraphics[width=0.9\linewidth]{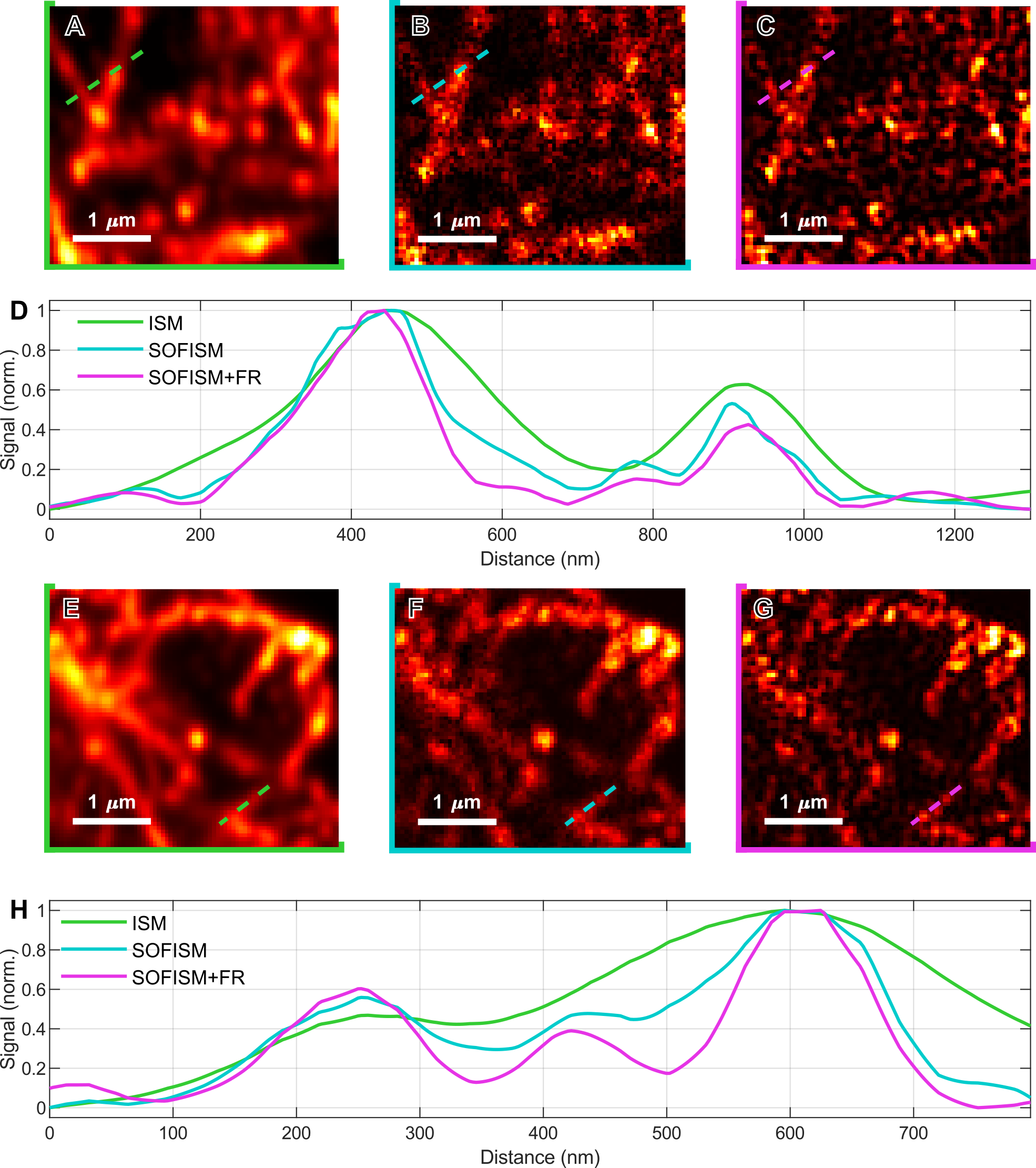}
    \caption{Comparison of ISM, SOFISM, and Fourier-reweighted SOFISM images of HeLa cells with CF568 and AF555 labeled microtubules (a--c) Representative ISM, SOFISM, and FR-SOFISM image for CF568, with the corresponding normalized intensity cross-section shown in (d). (e--g) Equivalent image for AF555, with the corresponding cross-section shown in (h).}
    \label{fig:fig5_emitters}
\end{figure}

Although AF647 provided the most robust performance, Fig.~\ref{fig:fig5_emitters} shows that the approach is not restricted to a single fluorophore. Both CF568 and AF555 produced usable SOFISM images, demonstrating flexibility in emitter choice for CW-based measurements. The CF568 dataset was acquired with four repeats, a 50~nm step size, and \SI{3}{\micro\watt} excitation power, whereas the AF555 dataset was acquired with four repeats, the same 50~nm step size, and \SI{5}{\micro\watt} excitation power. Temporal bins of \SI{2.5}{\micro\second} were used in both cases, and the accumulated delay ranges were selected using the same mean-to-STD analysis across detector-pair subsets. This gave delay cutoffs of 9\SI{2.5}{\micro\second} for CF568 and \SI{140}{\micro\second} for AF555. Under these conditions, CF568 yielded a visibly noisier image, while AF555 provided a higher-SNR result, as seen in the image panels and corresponding cross-sections.

\FloatBarrier

\section{Discussion}
\label{sec:discussion}

The results show that CW SOFISM is feasible when the relevant fluorescence fluctuations are sampled on timescales matched to the scanning dwell time. Both the sparse-emitter measurements and the HeLa microtubule imaging data indicate that the useful correlation signal is dominated by microsecond-to-sub-millisecond dynamics. The SPAD array is central to this operating regime, since its timestamped readout allows for dynamic binning that allows microsecond-scale temporal binning within each dwell period. This enables the effective frame rate to be optimized after acquisition. Although excessively fine binning can photon-starve the individual frames, this limit was not reached within the tested range. Instead, the shortest tested binning time gave the most reproducible SOFISM contrast in the mean-to-STD analysis in Fig.~\ref{fig:fig3_staticVsScan0} (e), provided that a sufficient range of correlation delays was accumulated. It is important to mention the technical tradeoffs, where finer temporal binning comes at the cost of greater memory use and longer computation times. In that sense the practical performance of CW SOFISM depends on preserving rapid fluorescence fluctuations throughout both acquisition and processing. We note that the power-line interference observed here was specific to the electronic coupling of our custom galvanometer drive. In commercial systems with highly integrated scanning electronics, such electrical coupling is negligible, which would circumvent the need for high-pass filtering and allow the preservation of slower blinking dynamics.

The tri-exponential decay observed for static AF647 emitters suggests that several photophysical processes contribute to the measured fluctuations. The fastest component, on the order of tens of microseconds, is consistent with triplet-state dynamics, which are commonly regarded as an important gateway to the long-lived dark states exploited in STORM-like switching schemes~\cite{songInfluenceTripletExcited1996}. The slower components in the hundreds of microseconds are plausibly associated with photo-isomerization or related dark-state processes that have been reported for cyanine dyes~\cite{karlssonPhotoisomerizationCyanineDye2019}. The measurements show that these dynamics are directly accessible in the present SPAD-based configuration, whereas they would be strongly averaged in conventional camera-based acquisition. This direct access to the relevant fluctuation timescales is important for SOFISM because it links fluorophore kinetics to the choice of temporal binning and accumulated correlation delays. 

The composition experiments further suggest that the method is robust to variations in the switching buffer, provided that both the oxygen-scavenging system and the reducing agent are present. 
This finding is consistent with the role of MEA in facilitating fluorophore switching. In cyanine dyes such as AF647, photoexcitation enables MEA to react with the polymethine bridge, disrupting the conjugated \(\pi\)-electron system and driving the fluorophore into a reversible non-fluorescent dark state~\cite{yangElectrochemicallyControlledBlinking2024,lindeHowSwitchFluorophore2014}. This reversible transition provides the on/off dynamics required for the stochastic intensity fluctuations used to build the SOFI/SOFISM contrast.

This is encouraging from a practical standpoint, because it indicates that usable CW SOFISM contrast does not rely on an exceptionally narrow sample preparation parameter window. At the same time, the repeated scans show that photobleaching is still present and progressively reduces the correlation amplitude over the course of an acquisition. The present buffer conditions therefore mitigate, but do not eliminate, the trade-off between maintaining fluorescence fluctuations and preserving signal over repeated scans. In practice, excitation power, dwell time, and number of repeats must still be balanced against bleaching when choosing imaging conditions.

Finally, measurements with CF568 and AF555 demonstrate that the method is not restricted to AF647, although the fluorophores did not perform equally well under the CW excitation conditions tested here. This difference is expected, since the detected SOFISM signal depends not only on labeling density and optical alignment, but also on the photophysical properties of the emitter, including its absorption cross section, fluorescence quantum yield, photostability, and, critically, the amplitude and timescale of its intensity fluctuations. AF647 combines a large absorption cross section with well-established blinking behavior in reducing/oxygen-scavenging buffers, and therefore produces the most robust and highest-contrast SOFISM images in these measurements. In comparison, CF568 and AF555 produced weaker but still resolvable fluctuation-based contrast, demonstrating that CW SOFISM can be extended to additional labeling schemes when the fluorophore provides sufficient brightness and suitable blinking dynamics. More broadly, these results support the use of fluctuation-based super-resolution in scanned ISM-type architectures without requiring pulsed excitation.

\section{Conclusion}
\label{sec:conclusion}
This work establishes SOFISM on a CW-equipped confocal system as a practical fluctuation-based super-resolution approach. The timing resolution of the SPAD array enabled correlation measurements with microsecond-scale temporal binning, which made it possible to resolve the fast fluorescence dynamics required for SOFISM under confocal conditions.
The switching-buffer conditions were sufficient to obtain a measurable correlation signal from AF647, CF568, and AF555 while promoting emitter fluctuations and limiting photobleaching enough for constructing an image. Within the tested range, variations in buffer composition had only a very minor effect on the extracted bi-exponential correlation times and on the final image contrast, indicating that the method is reasonably robust provided that both the oxygen-scavenging and reducing components are present.
The mean-to-STD analysis provided a practical criterion for selecting the frame rate and accumulated delay range that produced the most stable SOFISM contrast. The sparse-emitter sample measurements showed that substantial excitation power is required to drive the fluorophores into a fluctuation regime compatible with SOFISM on a CW system, with observed dynamics consistent with triplet-state or photo-isomerization-related blinking.
More broadly, the method requires only modest modifications to a confocal image-scanning architecture equipped with a pixelated SPAD detector. In that sense, it remains close in spirit to commercially available image-scanning implementations, while extending them with fluctuation-based super-resolution through suitable sample preparation and time-resolved data analysis. These results support SOFISM with CW excitation as an experimentally accessible route to improved spatial resolution in scanned fluorescence microscopy without resorting to pulsed excitation.

\section*{Funding}
This work was supported by the Israel Science Foundation (grant No. 1249/25) and by the Israeli Ministry of Science under the joint Israel-UK program. 
DO is the incumbent of the Harry Weinrebe Professorial Chair of Laser Physics.

\section*{Acknowledgments}

The authors thank U. Rossman for helpful discussions and his contribution
to the early development and construction of the experimental setup. The authors
also thank D. Nakar for helpful discussions and thoughtful suggestions regarding
figure design.

\section*{Disclosures}
The authors declare no conflicts of interest.

\section*{Data Availability Statement}
The MATLAB code used for the SOFISM analysis pipeline is available at: \url{https://github.com/Liorbeck/SOFISM-CW_Project}. The repository contains the main live script and modular source code for timestamp binning, sparse data generation, SOFISM correlation analysis, stability-based delay selection, measurement-dependent pixel reassignment, and Fourier reweighting. The example datasets used with this pipeline are available separately through Zenodo at \url{https://doi.org/10.5281/zenodo.20542920}. The data repository contains the microscopy datasets required to run the example analysis. Instructions for downloading the data and running the analysis are provided in the repository README file.
Additional information on the data format or analysis pipeline is
available from the corresponding author upon reasonable request.

\bibliographystyle{unsrtnat}
\bibliography{sofism_buffer_paper}

\appendix
\section{Supplementary methods and controls}
\label{sec:appendix}
\setcounter{figure}{0}
\setcounter{table}{0}
\setcounter{equation}{0}
\renewcommand{\thefigure}{S\arabic{figure}}
\renewcommand{\thetable}{S\arabic{table}}
\renewcommand{\theequation}{S\arabic{equation}}

\renewcommand{\theHfigure}{S\arabic{figure}}
\renewcommand{\theHtable}{S\arabic{table}}
\renewcommand{\theHequation}{S\arabic{equation}}

\subsection{Sample preparation and imaging buffers}
\label{sec:si_sample_prep_buffers}
\subsubsection*{Sparse AF647 samples}

Sparse AF647 samples were prepared by nonspecific adsorption of goat anti-mouse IgG secondary antibody AF647-conjugated (A21235, Invitrogen) onto poly-L-lysine-coated glass. Glass coverslips were
incubated with \SI{1}{mg/ml} poly-L-lysine solution (Sigma-Aldrich, P2636) for
\SI{15}{\minute} and dried before use. AF647-conjugated secondary antibody was then
applied to the coated glass and incubated for \SI{2}{\minute}. The sample was washed
to remove unbound antibody and mounted in imaging buffer.

\subsubsection*{HeLa cell culture}

HeLa cells were cultured in DMEM supplemented with \SI{10}{\percent} fetal bovine serum (FBS),
\SI{2}{mM} glutamine,
\SI[group-separator={,}, group-minimum-digits=4]{10000}{U/ml} penicillin, and
\SI{10}{mg/ml} streptomycin. Cells were maintained at
\SI{37}{\degreeCelsius} in a humidified atmosphere containing \SI{5}{\percent} CO\textsubscript{2} and
plated at low density on \SI{35}{mm} glass-bottom \#1.5 dishes (81218, Ibidi).

\subsubsection*{Immunofluorescence staining}

After \SI{24}{h}, cells were washed with warm PBS and fixed with warm \SI{4}{\percent}
paraformaldehyde for \SI{15}{min}. The cells were then permeabilized with
\SI{0.5}{\percent} Triton X-100 for \SI{10}{min} and blocked with \SI{10}{\percent} FBS for
\SI{15}{min}.

Cells were incubated overnight at \SI{4}{\degreeCelsius} with monoclonal
anti-$\alpha$-tubulin primary antibody (DM1A, Sigma-Aldrich). Secondary-antibody
labeling was performed for \SI{1}{h} at room temperature using CF568 anti-mouse
secondary antibody (SAB4600082, Sigma-Aldrich), or Alexa Fluor 555 and Alexa
Fluor 647 anti-mouse secondary antibodies (Thermo Fisher Scientific). Following
secondary-antibody incubation, cells were washed with PBS and the medium was
replaced with imaging buffer.

\subsubsection*{Imaging buffer preparation}
Stock solutions: Buffer B consisted of \SI{50}{mM} Tris-HCl, \SI{10}{mM} NaCl, and
\SI{10}{\percent} (w/v) glucose at pH~8.0. A GLOXY stock solution was prepared
by adding glucose oxidase and catalase to \SI{50}{mM} Tris and \SI{10}{mM}
NaCl at pH~8.0, with final activities of \SI{8440}{AU} glucose oxidase and
\SI[group-separator={,}, group-minimum-digits=4]{70200}{AU} catalase per
\SI{1}{mL} stock solution. A \SI{1}{M} MEA stock solution was prepared
separately.

Immediately before imaging, GLOXY stock solution and MEA stock solution were
mixed with Buffer B to prepare \SI{1}{mL} of imaging buffer. The final buffer
composition varied between measurements, as described in the relevant figure
captions and tables. When reported as percentages, MEA and GLOXY denote
stock-solution fractions in the final imaging-buffer volume.

\subsection{Data acquisition and detector configuration}

Data acquisition and analysis were performed using custom MATLAB routines. The
SPAD cameras were operated in timestamped streaming mode, allowing the photon
event buffer to be read continuously during acquisition and preventing buffer
overflow during long scans.
An external pulse generator provided the clock signal for the SPAD timestamping
electronics at \SI{1}{MHz}. Scan synchronization signals were generated by the
DAQ unit using TTL pulses. These pulses marked the pixel dwell timing and line
transitions, allowing the photon timestamps to be mapped onto the scan grid.
Each scan repetition was saved as a separate acquisition and combined during
post-processing.
Fluorescence from the \SI{640}{nm} and \SI{532}{nm} excitation channels was separated in the
detection path using a dichroic mirror and directed onto two independent SPAD
array cameras (SPAD23G, Pi Imaging). The two detection arms were equipped with
separate magnification optics, allowing the fluorescence spot size on each SPAD
array to be adjusted independently. This was required because the diffraction-limited
emission spot size depends on wavelength, and therefore the optimal
projection onto the detector array differs between the two spectral channels.



\FloatBarrier
\subsection{Timestamp binning and SOFISM analysis}
\label{sec:si_sofism_pipeline}

\begin{figure}[!htbp]
    \centering
    \includegraphics[width=0.5\linewidth]{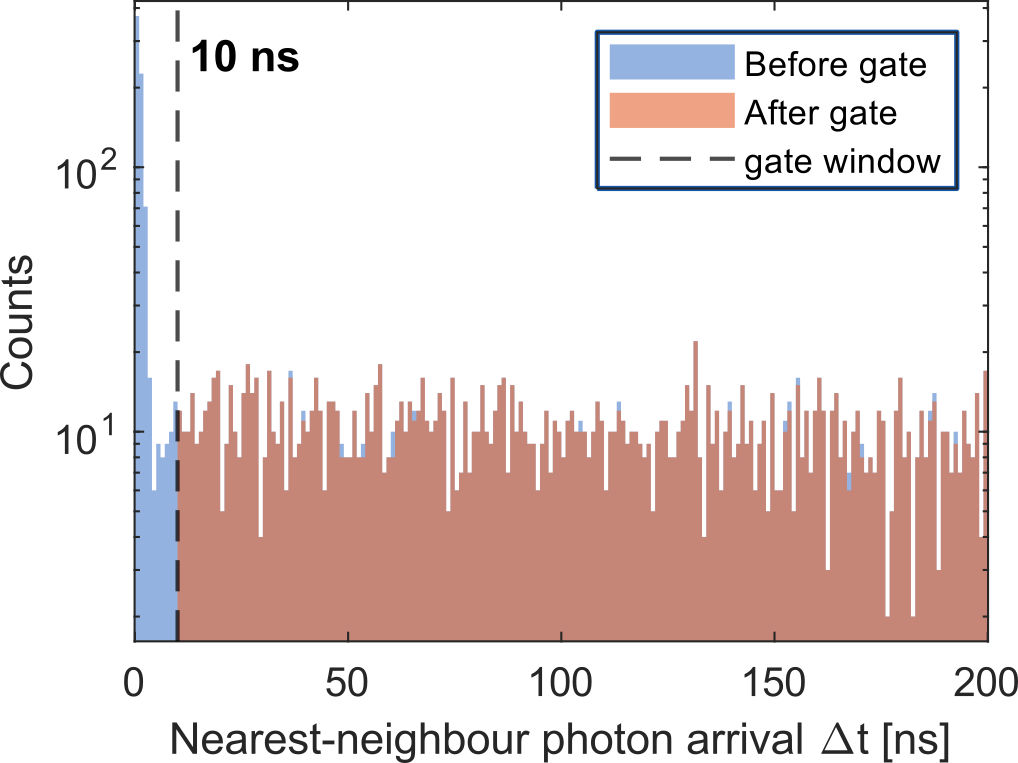}
    \caption{Nearest-neighbour photon-gap histogram before and after cross-talk
rejection.}
    \label{fig:SI_CrossTalkRemovalplot}
\end{figure}

The SOFISM analysis started from the raw timestamped photon stream recorded by
the SPAD array. Each event contained a marker identifying the event type, a
coarse clock value, and a fine time-to-digital converter (TDC) value. After
correcting for coarse-counter wraps, the photon arrival time was reconstructed as
\begin{equation}
  t_n = c_n T_{\mathrm{coarse}} + f_n
  \label{eq:abs_time}
\end{equation}
where $c_n$ is the corrected coarse-clock count, $T_{\mathrm{coarse}}$ is the
coarse-clock period, and $f_n$ is the fine TDC time.
The timestamp stream was sorted in time and mapped onto the scan grid using the
line and pixel synchronization markers together with the known scan coordinates.
For each scan
position, the remaining photons were binned into temporal bins of width
\(\Delta t_{\mathrm{bin}} = \SI{2.5}{\micro\second}\) and assigned to their
corresponding detector element. The resulting data can be regarded as a
five-dimensional intensity array with coordinates
\[
(y, x, t, i, r)
\]
where $(y, x)$ is the scan position, $t$ is the temporal bin within the pixel
dwell time, $i$ is the detector index, and $r$ is the scan repeat. In practice,
the data were stored in a sparse representation, retaining only nonzero photon
counts, to reduce memory usage during processing.
Detector cross-talk was mitigated by applying a short anti-coincidence gate to the
time-ordered photon stream. Pairs or clusters of photons detected on different
SPAD elements within a time window \(\Delta t_{\mathrm{xt}} = \SI{10}{ns}\) were
rejected before the sparse representation was assembled. This window was chosen
from the nearest-neighbour photon-gap histogram, where cross-talk events
appeared as a narrow sub- 10~ns peak above the accidental-coincidence
background as seen in Fig.~\ref{fig:SI_CrossTalkRemovalplot}.

Before correlation analysis, the temporal traces were trimmed at the beginning
and end of each pixel dwell time to remove intervals affected by galvo settling
and dwell-window edge effects. \(\SI{300}{\micro\second}\) were removed from the
beginning of each dwell window and \(\SI{100}{\micro\second}\) from the end.
The remaining traces were high-pass filtered along the temporal axis to suppress
a pronounced \SI{100}{Hz} intensity modulation originating from the galvo
scanning system. The filter was applied in the
Fourier domain using a smooth sigmoidal transfer function,
\begin{equation}
  H(f) =
  \frac{1}{1 + \exp[-k(f-f_c)]}
  \quad|\quad  k = \frac{4}{w}
  \label{eq:sigmoid_hp}
\end{equation}
with cutoff frequency \(f_c = \SI{250}{Hz}\) and transition width
\(w = \SI{20}{Hz}\), the DC component was set to zero. To
reduce ringing from the finite trace length, each trace was mirror-padded before
Fourier filtering and cropped back to its original length after the inverse
transform~\cite{panDesignWindowlessDigital2018}.
For each scan position and detector pair, fluctuation cross-correlations were
calculated from the filtered traces. For detector elements $i$ and $j$, the
correlation at delay $\tau$ was computed as
\begin{equation}
  C_{ij}(\mathbf{r}, \tau)
  =
  \frac{1}{N_{\mathrm{t}}-\tau}
  \sum_{m=1}^{N_{\mathrm{t}}-\tau}
  \delta I_i(\mathbf{r}, m)
  \delta I_j(\mathbf{r}, m+\tau)
  \label{eq:si_sofi_corr}
\end{equation}
where $\delta I_i(\mathbf{r}, m)$ is the filtered, mean-subtracted intensity
trace of detector $i$ at scan position $\mathbf{r}$, and $N_{\mathrm{t}}$ is the number of
valid temporal bins after trimming. Correlations were evaluated up to a maximum
delay $\tau_{\mathrm{max}}$.

A dead-time correction was applied to compensate for photon losses during the
SPAD recovery interval. The detectors were modeled as non-paralyzable, for
which the measured count rate $m_i$ of detector $i$ underestimates the true
rate by a factor $(1 - m_i \tau_{\mathrm{dead}})$, where
$\tau_{\mathrm{dead}} = \SI{50}{ns}$ is the detector dead time. Since each
SOFISM term is a product of two detector intensities, the corresponding
detector-pair correlation was rescaled as
\begin{equation}
  C_{ij}^{\mathrm{corr}}(\mathbf{r}, \tau)
  =
  \frac{C_{ij}(\mathbf{r}, \tau)}
       {\bigl(1 - m_i \tau_{\mathrm{dead}}\bigr)\bigl(1 - m_j \tau_{\mathrm{dead}}\bigr)}
  \label{eq:si_deadtime}
\end{equation}
where $m_i$ was estimated per scan position from the mean count rate of
detector $i$ in that pixel. The correction was applied to the correlation
images of Eq.~\eqref{eq:si_sofi_corr} prior to pixel reassignment.
Autocorrelation ($i=j$) terms were excluded from the final SOFISM
image. The autocorrelations are affected both by detector afterpulsing and by the
distortion that the temporal high-pass filter introduces due to the zero-lag
correlation, neither of which contaminates the cross-detector terms. Specifically, because the sigmoidal high-pass filter removes the low-frequency DC baseline, the fluctuations are forced to average to zero, causing the filtered correlation (both auto- and cross- correlations) to cross below zero and exhibit negative side lobes at non-zero delays. Furthermore, the strong shot-noise peak at zero delay ($\tau=0$) in the autocorrelation forces the rest of the delays to be negative to compensate the high shot-noise and integrate to 0.
Corner detectors were excluded from the analysis, and one detector in the \SI{640}{nm}
channel was additionally excluded as a hot pixel.
For the static measurements of the sparse sample, temporal segmentation was
used, where each dwell trace was divided into shorter segments that were treated as
additional statistical realizations before averaging. Otherwise, each scan
repeat was processed as a single realization.
The detector-pair correlation images were then shifted according to the
measurement-dependent PR vectors described in
Sec.~\ref{sec:si_measurement_dependent_pr}. The final images were Fourier-reweighted as described in the main
text.
\FloatBarrier
\subsection{Measurement-dependent pixel reassignment}
\label{sec:si_measurement_dependent_pr}

\begin{figure}[!htbp]
    \centering
    \includegraphics[width=0.9\linewidth]{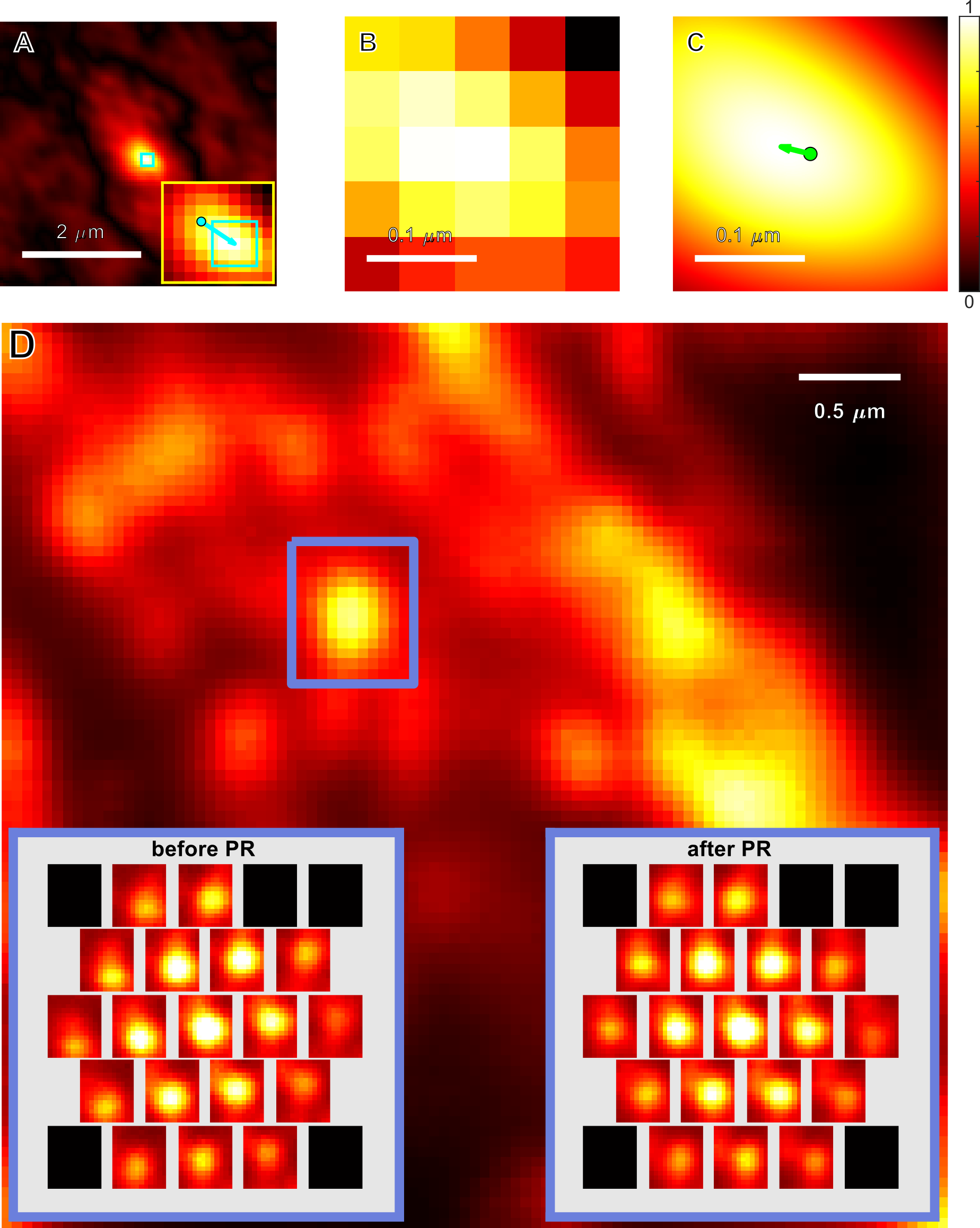}
    \caption{Measurement-dependent pixel reassignment.
    All panels were generated from the same HeLa microtubule measurement used
    to estimate the detector-dependent parallax shifts.
    \textbf{(a)}~Lateral cross-correlation between the reference detector
    and an off-center detector, computed from their mean-subtracted
    gradient-magnitude images. The peak is offset from the origin by the
    parallax shift between the two detectors. The cyan arrow marks the
    integer-pixel displacement and the cyan box marks the \(\pm 2\)~pixel
    neighborhood refined in~(b, c).
    \textbf{(b)}~The \(\pm 2\)~pixel neighborhood around the integer-pixel
    peak.
    \textbf{(c)}~The cross-correlation in the same neighborhood, re-evaluated
    on a finely spaced grid of candidate displacements (\num{0.01}-pixel
    spacing). The green arrow marks the sub-pixel correction to the integer
    estimate, giving the final parallax shift.
    \textbf{(d)}~Confocal (CLSM) image of HeLa microtubules; insets show a
    magnified region rendered on the 23-element hexagonal detector grid,
    before pixel reassignment (left) and after pixel reassignment (right),
    illustrating the alignment improvement.
    }
    \label{fig:SI_PRplot}
\end{figure}

Pixel reassignment was performed by measuring the apparent lateral displacement of the scanned image on each detector element. Because each detector samples a different position in the image plane, the same emitter appears with a detector-dependent parallax shift. These shifts were estimated directly from the measured detector images, using the central detector as the reference detector.
For each detector element, the sum of the scanned image was compared to the image recorded by the central detector. Before estimating the displacement, the raw intensity images were converted to gradient-magnitude images using the \texttt{imgradient} function in MATLAB. The lateral derivatives of the image were estimated this way for each detector on the pixelated SPAD camera.
\begin{equation}
    I_a^{\nabla}(x, y)
    =
    \left|\nabla I_a(x, y)\right|
    =
    \sqrt{
    \left(\frac{\partial I_a}{\partial x}\right)^2+
    \left(\frac{\partial I_a}{\partial y}\right)^2
    }
\end{equation}
The spatial mean of this gradient image was then subtracted:
\begin{equation}
    \tilde{I}_a^{\nabla}(x, y)
    =
    I_a^{\nabla}(x, y)
    -
    \left\langle I_a^{\nabla}(x, y)\right\rangle_{x, y}
\end{equation}
This step makes the shift estimation depend primarily on the image's spatial structure, such as emitter positions and image edges, rather than on the absolute brightness of each detector element.
To reduce edge artifacts in the Fourier calculation, the mean-subtracted gradient images were multiplied by a two-dimensional Tukey window with cosine fraction 0.2. The lateral translation between detector $a$ and the central detector was then estimated from their Fourier cross-correlation. First, the integer-pixel displacement was found from the maximum of the cross-correlation image. The integer-pixel estimate was then refined to sub-pixel precision. The discrete cross-correlation is a sampling of an underlying continuous function of the lateral displacement; the fast Fourier transform evaluates it only at integer-pixel lags, but the same Fourier sum can be evaluated at any real-valued displacement. The cross-correlation was therefore re-evaluated on a finely spaced grid of candidate displacements, with a spacing of \num{0.01} scan pixels, within a small neighborhood of the integer-pixel peak. For a fixed set of candidate displacements, this evaluation is linear in the cross-power spectrum and was computed efficiently as a pair of matrix multiplications with precomputed Fourier-kernel matrices~\cite{guizar-sicairosEfficientSubpixelImage2008}. The sub-pixel parallax shift was taken as the candidate displacement that maximized the lateral gradient cross-correlation. The process of finding the PR vectors is shown in Figure~\ref{fig:SI_PRplot} (a--c).
The measured parallax vector of detector $a$ relative to the central detector was denoted
\begin{equation}
    \vec{\nu}_a
    =
    \left(
    \Delta x_a, \Delta y_a
    \right)
\end{equation}
with
\begin{equation}
    \vec{\nu}_{\mathrm{center}} = (0, 0).
\end{equation}
Each detector image was then translated by $-\vec{\nu}_a$ using bicubic interpolation, so that the images from all detector elements were brought onto the same lateral coordinate system before summation. An example of the images before and after the PR procedure is shown in Figure~\ref{fig:SI_PRplot}(d).

For second-order SOFISM, the same single-detector parallax vectors were used to translate each detector-pair image. For a detector pair $(a, b)$, the required translation vector was taken as the mean of the two detector vectors, since the signal originates from the midpoint between the detectors:
\begin{equation}
    \vec{\nu}_{ab}
    =
    \frac{\vec{\nu}_a + \vec{\nu}_b}{2}
\end{equation}
More generally, for an $n$th-order cumulant image formed from detectors $a_1, \ldots, a_n$, the corresponding translation vector is
\begin{equation}
    \vec{\nu}_{a_1, \ldots, a_n}
    =
    \frac{1}{n}
    \sum_{j=1}^{n} \vec{\nu}_{a_j}
\end{equation}




\FloatBarrier
\subsection{Buffer behavior}
\label{sec:si_buffer_behavior}
\begin{figure}[!htbp]
    \centering
    \includegraphics[width=0.9\linewidth]{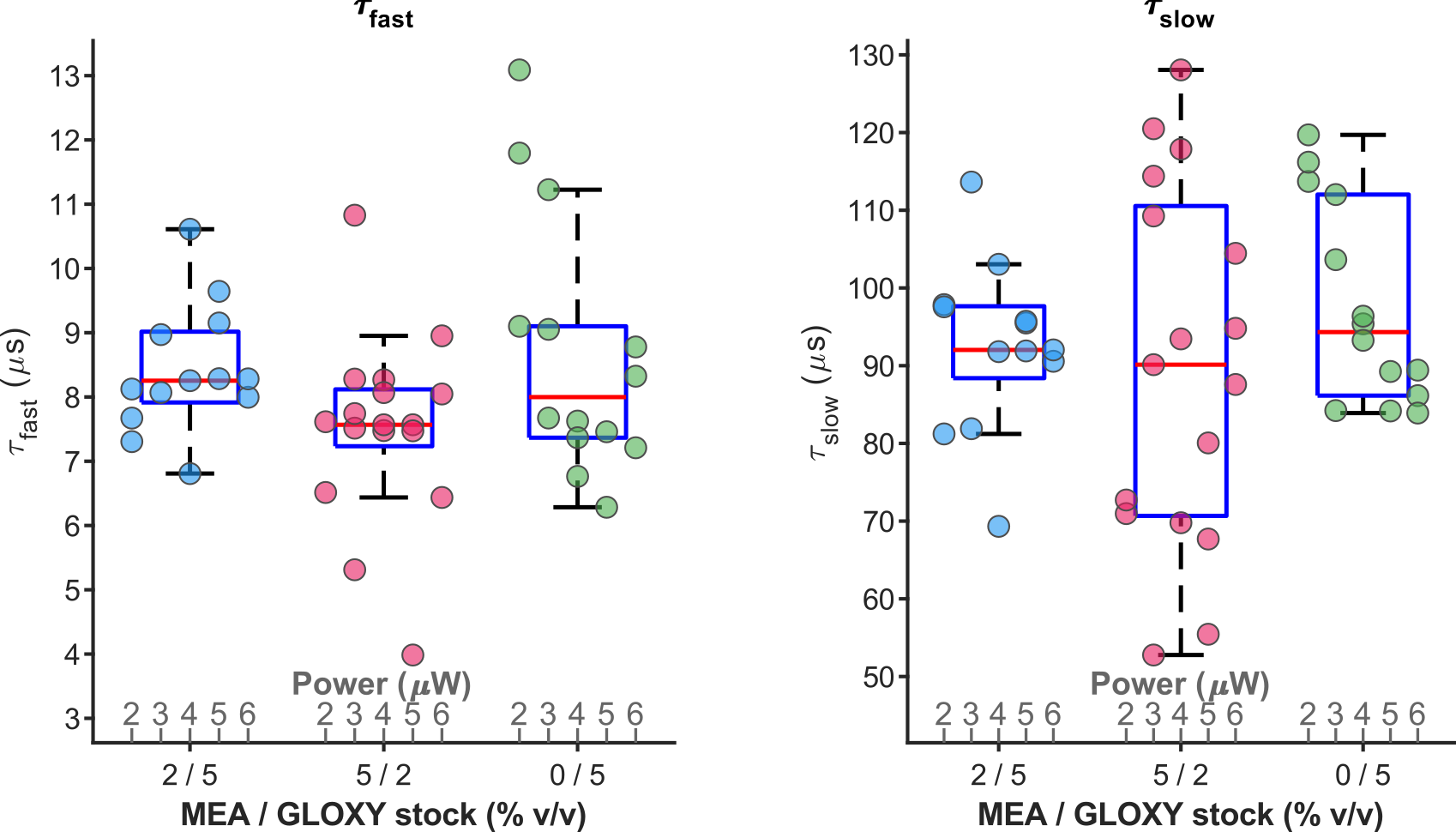}
    \caption{Distribution of fitted correlation time constants as a function of excitation power for the tested buffer conditions. The plot summarizes how the extracted decay times vary with illumination power and complements the buffer-composition comparison in the main text.}
    \label{fig:si_tau_boxplot}
\end{figure}

Multiple images of the HeLa microtubules stained with AF647 were captured at different illumination powers and buffer compositions, altering the amount of MEA and GLOXY. The total correlation results were summed over the image and fitted to a bi-exponential decay. The extracted decay times showed only small variability with either power or buffer composition, as summarized in Fig.~\ref{fig:si_tau_boxplot}.
The only exception is the buffer composition without MEA and with \SI{5}{\percent}~v/v GLOXY stock, where an approximately monotonic decrease was observed for the slow component, from about \SI{120}{\micro\second} to \SI{80}{\micro\second} as the excitation power increased from \SI{2}{\micro\watt} to \SI{6}{\micro\watt}.

\begin{figure}[!htbp]
    \centering
    \includegraphics[width=0.9\linewidth]{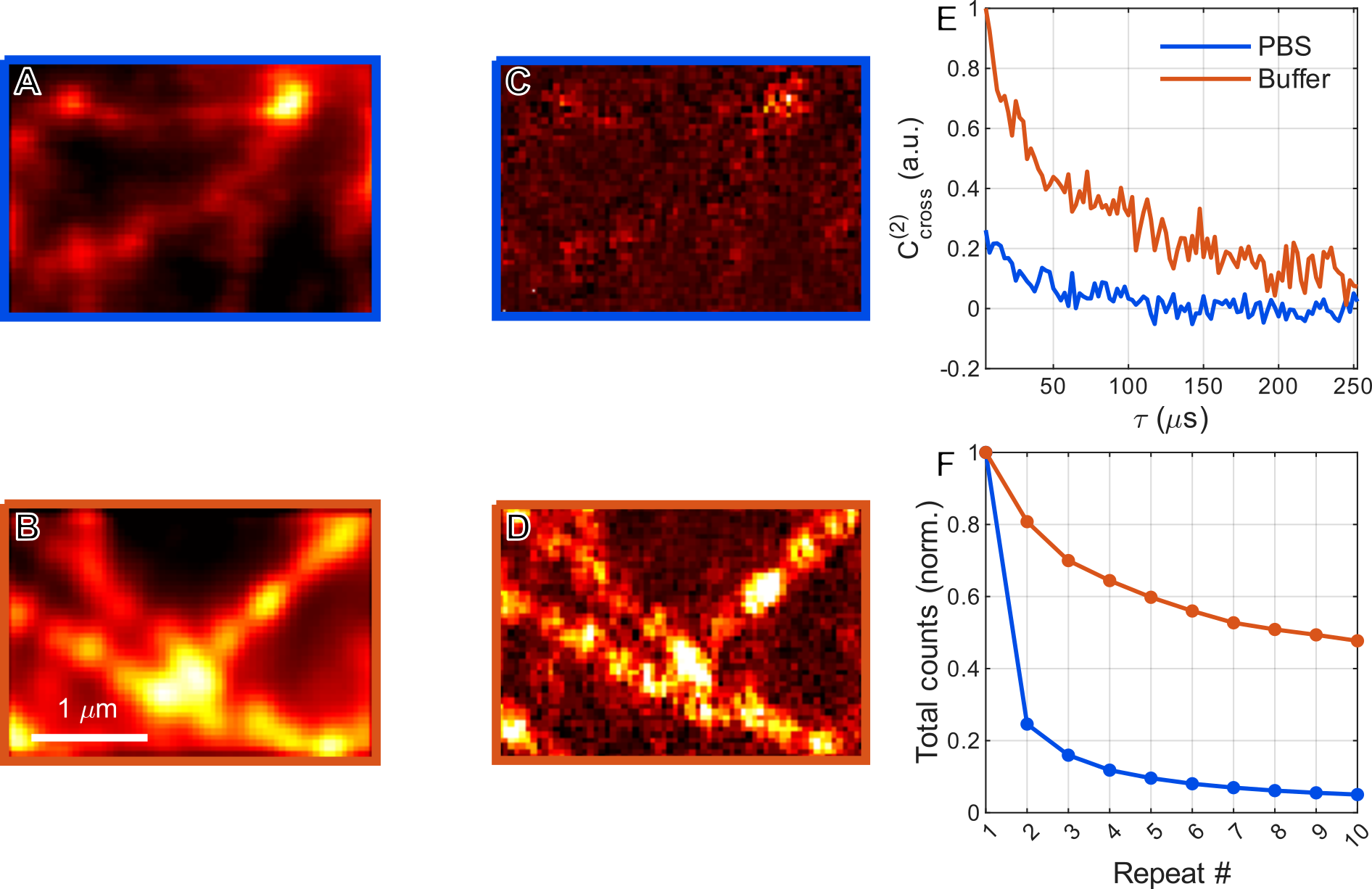}
    \caption{HeLa microtubule sample stained with AF647, \SI{10}{\milli\second}
    dwell time and \SI{50}{\nano\meter} step size.
    \textbf{(a)} ISM image in PBS, excitation \SI{3}{\micro\watt}, 10 repeats.
    \textbf{(b)} ISM image in 0/5 (\SI{5}{\percent}~v/v) MEA/GLOXY stock,
    excitation \SI{3}{\micro\watt}, 10 repeats.
    \textbf{(c, d)} Corresponding SOFISM cross-correlation images for (a) and (b),
    summed over delays $\tau$.
    \textbf{(e)} Cross-correlation $C^{(2)}_{\mathrm{cross}}(\tau)$, averaged over
    the ISM signal region defined by a per-dataset Otsu threshold, plotted
    against delay.
    \textbf{(f)} Total detected counts per repeat, normalized to the first
    repeat, showing photobleaching over the acquisition.}
    \label{fig:si_PBSCompFigure}
\end{figure}

To verify that the recovered cross-correlations require buffer-induced fluorescence fluctuations, AF647-labelled HeLa microtubules were imaged in plain PBS and in MEA/GLOXY imaging buffer as described above. Figure~\ref{fig:si_PBSCompFigure} shows an example of the 0/5 MEA/GLOXY condition compared against the PBS measurement. In both cases the ISM image shows clear microtubule structure, but a meaningful cross-correlation signal is recovered only in the switching buffer. In PBS the image structure is still present, yet the summed cross-correlation is too weak to give meaningful contrast in the SOFISM image, consistent with the absence of appreciable blinking and with rapid photobleaching under these conditions. Both datasets were acquired with a \SI{50}{\nano\meter} scan step size, 10 repeats, and \SI{3}{\micro\watt} excitation power.

\FloatBarrier
\subsection{Selection of frame rate and delay}
\label{sec:si_selection_framerate_delay}
The set of detector-pair correlations was divided into $K = 10$ disjoint groups. 
To avoid groups being dominated by different detector-pair brightness levels, the pairs were first ranked by their reference SOFISM contribution within the ISM feature mask and then divided between groups in a way that kept the brightness levels balanced. Autocorrelation terms were excluded along with the corner pixels and a hot pixel. 

For each tested maximum accumulated delay $d_{\mathrm{max}}$, a partial SOFISM contribution image was computed from each detector-pair group. Denoting the accumulated contribution from group $k$ by $G_k(\mathbf{r}, d_{\mathrm{max}})$, where $\mathbf{r}$ is the image coordinate, the group-averaged image was

\begin{equation}
\bar{G}(\mathbf{r}, d_{\mathrm{max}})
=
\frac{1}{K}
\sum_{k=1}^{K}
G_k(\mathbf{r}, d_{\mathrm{max}})
\end{equation}

The variability across detector-pair groups was estimated directly from the standard deviation of these group contribution images,

\begin{equation}
\sigma_G(\mathbf{r}, d_{\mathrm{max}})
=
\sqrt{
\frac{1}{K-1}
\sum_{k=1}^{K}
\left[
G_k(\mathbf{r}, d_{\mathrm{max}})
-
\bar{G}(\mathbf{r}, d_{\mathrm{max}})
\right]^2
}
\end{equation}

A local stability metric was then defined as

\begin{equation}
S_{\mathrm{pair}}(\mathbf{r}, d_{\mathrm{max}})
=
\frac{
\bar{G}(\mathbf{r}, d_{\mathrm{max}})
}{
\sigma_G(\mathbf{r}, d_{\mathrm{max}})
}
\end{equation}

\begin{figure}[!htbp]
    \centering
    \includegraphics{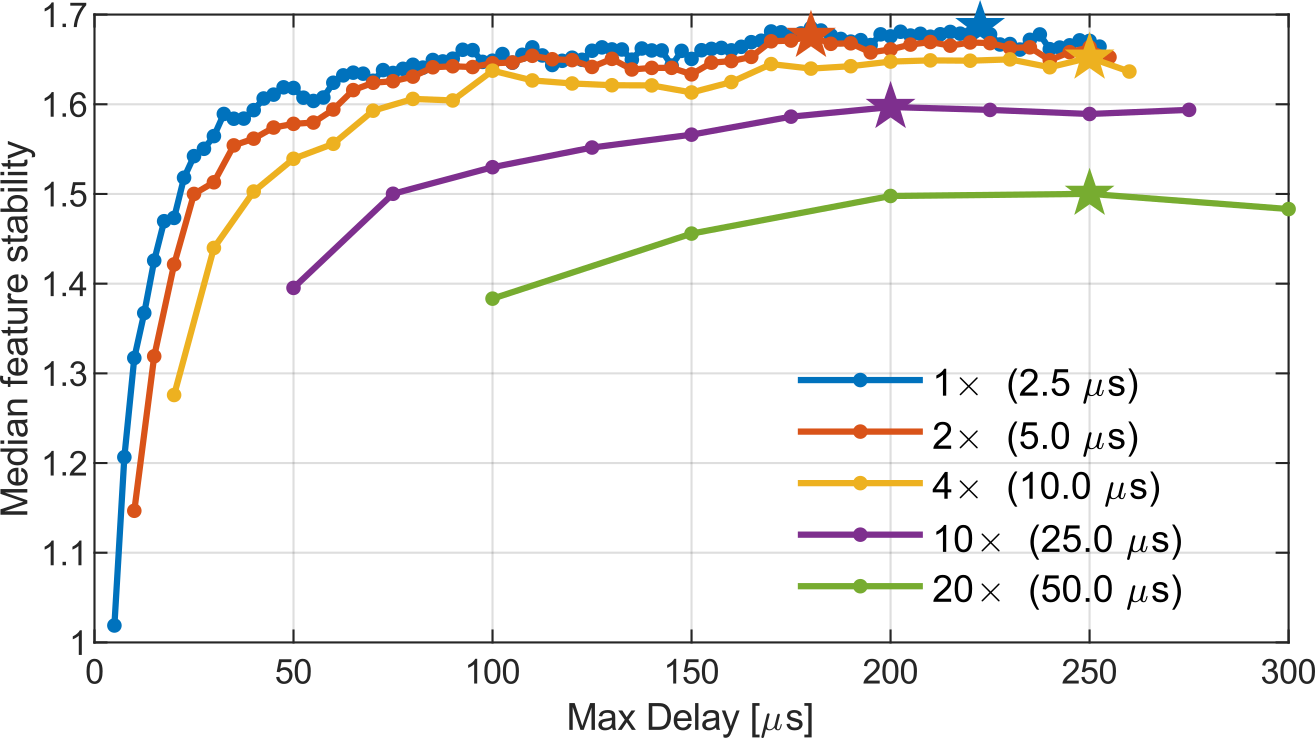}
    \caption{Repeat-based stability metric $S_{\mathrm{rep}}$ versus maximum
    accumulated correlation delay for several temporal binnings, evaluated
    across the five scan repetitions within the ISM feature mask. Stars mark
    the per-binning optimum.}
    \label{fig:SI_STDByRepeat}
\end{figure}

This quantity is an empirical measure of how reproducibly the SOFISM contrast is supported across balanced subsets of detector cross-correlation pairs. To avoid bias from background-dominated regions, the analysis was restricted to a feature mask derived from the corresponding ISM image using Otsu thresholding~\cite{otsuThresholdSelectionMethod1979}.
To confirm that the chosen binning does not depend on how the detector-pair
correlations were grouped, the same stability analysis was carried out using
the $R_{\mathrm{rep}} = 5$ successive scans of the field of view as independent
realizations, in place of the detector-pair groups. Each scan yields one
SOFISM contribution image, $G_r(\mathbf{r}, d_{\mathrm{max}})$, formed from the
full set of cross-correlation pairs across the detector array, so that every
realization shares the same confocal detection geometry and the same
pair-to-pair signal weighting. The stability metric is formed in the same way,
\begin{equation}
S_{\mathrm{rep}}(\mathbf{r}, d_{\mathrm{max}})
=
\frac{
\bar{G}(\mathbf{r}, d_{\mathrm{max}})
}{
\sigma_G(\mathbf{r}, d_{\mathrm{max}})
}
\end{equation}
with $\bar{G}$ and $\sigma_G$ now evaluated across the repeated scans, within
the same ISM feature mask.
Neither metric is conclusive on its own. The detector-pair metric groups
cross-correlation terms that are not mutually independent, since all detector
pairs collect light from the same blinking emitters within the confocal
detection volume. The repeat-based metric is limited by the small number of
scans, $R_{\mathrm{rep}} = 5$, and is biased low by photobleaching, the emitter
brightness decreases monotonically from one scan to the next, so part of
$\sigma_G$ reflects this bleaching rather than the fluctuations of interest.
Both metrics nonetheless show the same trend, shorter pixel dwell binning
giving a higher stability metric once enough correlation delays are accumulated
(Fig.~\ref{fig:SI_STDByRepeat}).

\FloatBarrier
\subsection{Resolution increase estimation}

\begin{figure}[!htbp]
    \centering
    \includegraphics[width=\linewidth]{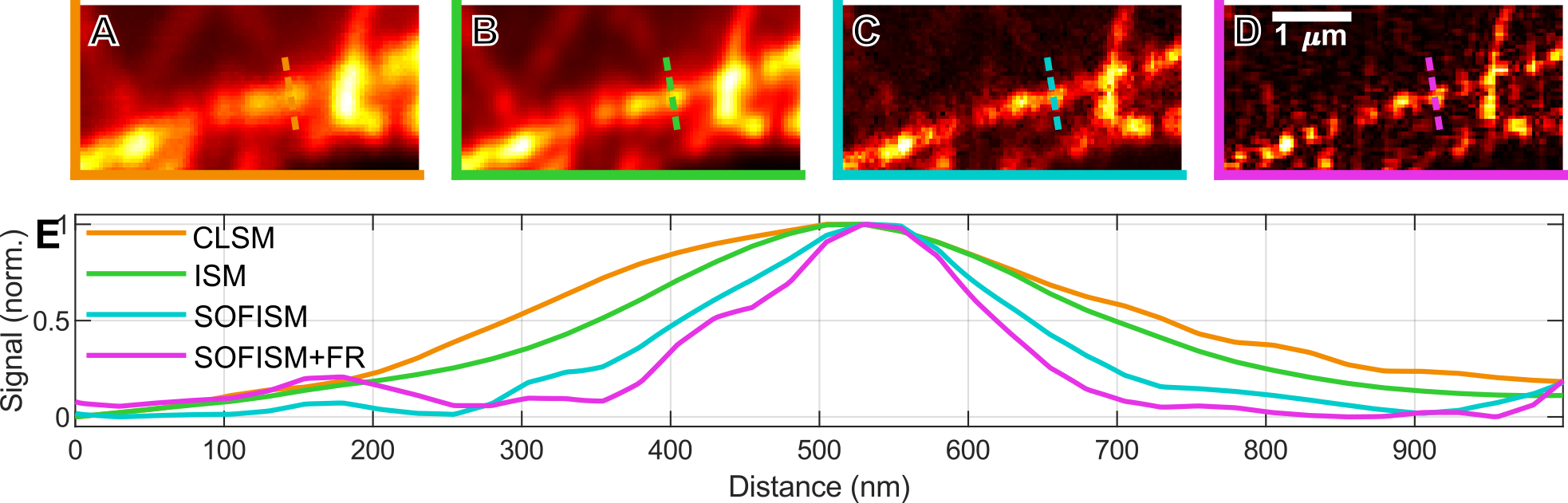}
    \caption{
    Image comparison from the same measurement, acquired with a \SI{10}{\milli\second} pixel dwell time, \SI{50}{\nano\meter} step size, 5 repetitions, and \SI{5}{\micro\watt} excitation power.
    (a) CLSM image, formed by summing the photons from all SPAD pixels.
    (b) ISM image.
    (c) SOFISM image.
    (d) Fourier-reweighted SOFISM image.
    (e) Cross-section across a microtubule feature, used to estimate the apparent feature width.
    }
    \label{fig:SI_dataSet5CombinedFigure}
\end{figure}

The expected resolution improvement factors can be understood by approximating the relevant effective point-spread functions as Gaussian. Under this approximation, multiplication of two similar Gaussian profiles narrows the width by a factor of $\sqrt{2}$. ISM alone can therefore improve the lateral resolution by up to a factor of $\sqrt{2}$ relative to a conventional confocal image. The second-order SOFI contribution can provide an additional factor of $\sqrt{2}$, giving an overall improvement of up to a factor of 2 for SOFISM. When Fourier reweighting is additionally applied, a further resolution improvement of up to $\sqrt{2}$ can in principle be obtained, corresponding to a total expected improvement of up to $2\sqrt{2}$.
The resolution improvement was estimated from an intensity cross-section across a microtubule feature, shown in Fig.~\ref{fig:SI_dataSet5CombinedFigure}. The same region was compared for the CLSM, ISM, SOFISM, and Fourier-reweighted SOFISM images. For each image, the apparent feature width was estimated from the full width at half maximum (FWHM) of the cross-section. The improvement factor was calculated as the ratio of consecutive FWHM values, relative to the original CLSM image. The values in~\ref{tab:SI_resolution_increase} align with the theory discussed, as each method produces a result nearing $\sqrt{2}$.

\begin{table}[!htbp]
\centering
\caption{Estimated resolution improvement from the FWHM of the microtubule cross-section shown in Fig.~\ref{fig:SI_dataSet5CombinedFigure}.}
\label{tab:SI_resolution_increase}
\begin{tabular}{lccc}
\hline
Image & FWHM (nm) & Step gain & Total gain \\
\hline
CLSM        & 451 & --   & -- \\
ISM         & 342 & 1.32 & 1.32 \\
SOFISM      & 242 & 1.41 & 1.86 \\
SOFISM-FR   & 197 & 1.23 & 2.29 \\
\hline
\end{tabular}
\end{table}

\end{document}